
\magnification = 1200
\def\lapp{\hbox{$ {
\lower.40ex\hbox{$<$}
\atop \raise.20ex\hbox{$\sim$}
}
$}  }
\def\rapp{\hbox{$ {
\lower.40ex\hbox{$>$}
\atop \raise.20ex\hbox{$\sim$}
}
$}  }
\def\barre#1{{\not\mathrel #1}}
\def\krig#1{\vbox{\ialign{\hfil##\hfil\crcr
$\raise0.3pt\hbox{$\scriptstyle \circ$}$\crcr\noalign
{\kern-0.02pt\nointerlineskip}
$\displaystyle{#1}$\crcr}}}
\def\upar#1{\vbox{\ialign{\hfil##\hfil\crcr
$\raise0.3pt\hbox{$\scriptstyle \leftrightarrow$}$\crcr\noalign
{\kern-0.02pt\nointerlineskip}
$\displaystyle{#1}$\crcr}}}
\def\ular#1{\vbox{\ialign{\hfil##\hfil\crcr
$\raise0.3pt\hbox{$\scriptstyle \leftarrow$}$\crcr\noalign
{\kern-0.02pt\nointerlineskip}
$\displaystyle{#1}$\crcr}}}

\def\svec#1{\skew{-2}\vec#1}
\def\Tr{\,{\rm Tr }\,}

\def\g5{\gamma_5}

\def\lp1{{\cal L}_{\pi N}^{(1)}}
\def\lp2{{\cal L}_{\pi N}^{(2)}}
\def\lp3{{\cal L}_{\pi N}^{(3)}}

\topskip=0.40truein
\leftskip=0.18truein
\vsize=8.8truein
\hsize=6.5truein
\tolerance 10000
\hfuzz=20pt

\baselineskip 14pt plus 1pt minus 1pt
\pageno=0
\centerline{\bf THRESHOLD TWO--PION PHOTO-- AND ELECTROPRODUCTION:}
\centerline{\bf More neutrals than expected}
\vskip 48pt
\centerline{V. Bernard$^1$, N. Kaiser$^2$,
Ulf-G. Mei{\ss}ner$^1$, A. Schmidt$^2$}
\vskip 16pt
\centerline{$^1${\it Centre de Recherches Nucl\'{e}aires et Universit\'{e}
Louis Pasteur de Strasbourg}}
\centerline{\it Physique Th\'{e}orique,
BP 20Cr, 67037 Strasbourg Cedex 2, France}
\vskip  4pt
\centerline{$^2${\it Physik Department T30,
Technische Universit\"at M\"unchen}}
\centerline{\it
James
Franck Stra{\ss}e,
D-85747 Garching, Germany}
\vskip  4pt
\vskip 12pt
\vskip 1.0in
\centerline{\bf ABSTRACT}
\medskip
\noindent We present an exploratory study of two pion photo-- and
electroproduction
off the nucleon in the threshold region. To calculate the pertinent amplitudes,
we make use of  heavy baryon chiral perturbation theory. We show that due to
finite chiral loops the production cross section for final states with two
neutral pions is considerably enhanced. The experimental implications are
briefly discussed.
\medskip
\vfill
\noindent CRN 94/14 \hfill March  1994
\vskip 12pt
\eject
\baselineskip 14pt plus 1pt minus 1pt
\noindent{\bf I. INTRODUCTION}
\bigskip
Over the last few years, much interest has been focused on pion photo-- and
electroproduction off nucleons. In particular, new accurate data for
the processes $\gamma p \to \pi^0 p$ and $\gamma^\star p \to \pi^0 p$ close to
production threshold have become available [1,2].\footnote{$^*$}{Here, $\gamma$
and $\gamma^\star$ stand for the real and virtual photon, respectively. In what
follows, we will exclusively use the symbol $\gamma$ but always spell out the
pertinent four--momentum
squared which is zero for real and less than zero for virtual
photons, in order.} These have led to many theoretical investigations. In
particular, in refs.[3] baryon chiral perturbation theory was used to give a
model--independent description of the pertinent differential cross sections,
multipoles and so on. While the study of these reactions,
also with charged pions in the final state, continues on the
experimental as well as on the theoretical side, complementary information
can be gained from the two pion production process $\gamma N \to \pi \pi N$,
where $N$ denotes the nucleon and $\gamma$ the real or virtual photons. The two
pions in the final state can both be charged, both neutral or one charged and
one  neutral. If the four--momentum of the photon is denoted by $k$, in case of
$k^2 = 0$ one speaks of photoproduction and for $k^2 < 0$ of electroproduction.
Here, we will be concerned with the threshold region, i.e. the photon in the
initial state has just enough energy to produce the two pions (and the outgoing
nucleon) at rest. This energy is very close to the first strong resonance
excitation of the nucleon, the $\Delta (1232)$ resonance. In fact, presently
available data focus on the resonance region and above. In that case, a
two--step reaction mechanism of the form $\gamma N \to \pi \Delta \to \pi \pi
N$ is appropriate to describe these data as detailed in refs.[4,5]. As we will
show later, there is, however, a narrow window above threshold which is
particularly sensitive to chiral loops, i.e. to the strictures of the
spontaneously broken chiral symmetry. In the limit of vanishing current quark
masses, which is a good first approximation for the two light flavors $u$ and
$d$  relevant to the $\pi N$ system, the gauge theory of the strong
interactions, QCD, exhibits this symmetry. While QCD is formulated in terms of
(confined) quark and gluon fields, its low--energy behaviour is dominated by
the
almost massless Goldstone bosons related to the spontaneous breakdown of the
chiral symmetry. This feature allows to formulate an effective field theory in
terms of the asymptotically observed fields (here, the pions and the nucleons)
which is amenable to a systematic energy expansion, baryon chiral perturbation
theory. This framework has been applied to a wide variety of hadronic and
nuclear processes as detailed in the reviews [6,7]. In the present context,
CHPT  amounts to a calculation of tree and pion one--loop diagrams.
The latter are mandated to perturbatively restore unitarity and they are also
demanded by the power counting scheme [8] underlying the effective field
theory. In the baryon sector, a consistent one--to--one mapping between the
energy and loop expansion is only possible if one transforms the nucleon mass
term into a string of vertices of increasing powers in the inverse nucleon
mass. This can be achieved in the extreme non--relativistic limit [9].
Evidently, such a procedure is nothing but a series of Foldy--Wouthuysen
transformations known from the treatment of the Dirac equation for a very heavy
fermion.
\medskip
What makes the process $\gamma N \to \pi \pi N$ so interesting is that, as we
will demonstrate in detail later on, the one--loop corrections lead to a
dramatic increase in the production rates for final states containing neutral
pions. This is a counter--intuitive result since in the case of single pion
production the cross sections for production of charged pions are considerably
larger than the ones with neutral pions in the final state. First
measurements of two--pion production at low energies have been performed at
MAMI
and we expect that the theoretical predictions discussed below will give
additional motivation to perform yet more detailed measurements of this
particular reaction. Our study extends the one of Dahm and Drechsel [10] who
discussed certain aspects of two--pion photoproduction in the framework of
Weinberg's chiral pion--nucleon Lagrangian [11].
\medskip
The paper is organized as follows. In section 2, we  discuss the
formal aspects of threshold two--pion production, such as kinematics,
threshold  multipoles
 and the three--body phase space. We also give formulae for the total
cross sections for the various final states, in particular we write down some
very handy approximations which allow the interested reader
to get a quick and fairly
accurate first estimate. We extend this discussion to the amplitudes close to
but not at threshold since these are important in the chiral expansion with two
charged pions in the final state (see appendix A).
Section 3 is devoted to a short discussion of
the effective chiral pion--nucleon Lagrangian. We also discuss briefly the
$\gamma \pi \Delta N$ system. The latter has to be included due
to the closeness of the $\Delta (1232)$ resonance. In section 4, the chiral
expansion of the threshold multipole amplitudes is given and the various
contributions are discussed. While the tree and one--loop diagrams
in the pion--nucleon system  lead to unambiguous results, the tree diagrams
including the $\Delta$ have potentially large uncertainties due to certain
badly known off--shell parameters. However, to the first two orders in
the chiral expansion only one of these enters and the final results are found
to be weakly dependent on it. Section 5 contains the numerical results.
We discuss the threshold multipoles and total cross sections making the
approximation that the  amplitude exactly at threshold represents the
 amplitude in the threshold region. For the particular channels $\gamma p \to
\pi^+ \pi^- p$ and $\gamma p \to \pi^0 \pi^0 p$,
we extend this calculation to the first non--vanishing
corrections away from threshold. The experimental implications are also
discussed. Most of the results presented are for the photoproduction case since
the longitudinal multipoles appearing in electroproduction do not carry any
relevant additional information to the first two orders in the chiral
expansion.
Section 6 contains the summary and outlook. Some technicalities
are relegated to appendix B.
\goodbreak \bigskip
\noindent{\bf II. TWO-PION PRODUCTION: FORMAL ASPECTS}
\bigskip
In this section, we will give the formalism necessary to treat two-pion photo-
and electroproduction in the threshold region. We will only be concerned with
the kinematics close or at threshold, the corresponding multipole decomposition
and the total cross sections. For a more general discussion we refer the reader
to [10].
\medskip
\noindent{\bf II.1. GENERAL CONSIDERATIONS}
\medskip
Consider the process $\gamma(k) + N(p_1) \to \pi^a(q_1) + \pi^b(q_2) + N(p_2)$,
with $N$ denoting the nucleon (proton or neutron), $\gamma$ a real ($k^2= 0$)
or virtual ($k^2 < 0$) photon and pions of isospin $a,b$. The pertinent
four-momenta are given (cf. fig.1) and the polarization vector of the photon is
denoted by $\epsilon_\mu$. The corresponding current transition matrix element
is
$$T\cdot \epsilon = \quad _{\rm out}<\pi^a(q_1),\pi^b(q_2),N(p_2)|
J^{em}_\mu(0)
\epsilon^\mu| N(p_1)>_{\rm in} \eqno(2.1)$$
with $J^{em}_\mu$ the electromagnetic current operator. From the two initial
states $\gamma p$ and $\gamma n$ we can form in total six final states
$$\eqalign{ \gamma p & \to \pi^+ \pi^- p\,, \cr \gamma p  & \to \pi^+ \pi^0 n
\,, \cr \gamma p & \to \pi^0 \pi^0 p\,, \cr \gamma n & \to \pi^+ \pi^- n\,,
\cr  \gamma n & \to \pi^0 \pi^- p\,, \cr \gamma n & \to \pi^0 \pi^0 n\,. \cr}
\eqno(2.2)$$
In what follows we will mostly concentrate on the first three of these
channels. To first order in the electromagnetic coupling $e$ the threshold
amplitudes for $\gamma p \to \pi^+ \pi^0 n$ and $\gamma n \to \pi^- \pi^0 p$
are equal as will become clear soon.
\medskip
\noindent{\bf II.2. KINEMATICS IN THE THRESHOLD REGION}
\medskip
In general, one can form five/six Mandelstam variables for the two-pion
photo/ electroproduction process from the independent four-momenta. For our
purpose, it is most convenient to work in the photon-nucleon center-of-mass
frame. At threshold, the real or virtual photon has just enough energy to
produce the two pions  at rest. Let us denote by $M_\pi$ and
$m$ the masses of the pion and nucleon, respectively.\footnote{*}{In section
II.5 we briefly discuss the relevance of differentiating the various masses in
the six processes (2.2).} The threshold center-of-mass energy squared is
$$s_{\rm thr} = (p_1 + k)^2_{\rm thr} =
(m + 2M_\pi)^2 = m^2 ( 1 + 4 \mu + 4 \mu^2)
\eqno(2.3)$$
where we have introduced the small parameter
$$\mu = { M_\pi \over m} \simeq {1\over 7}\,.\eqno(2.4)$$
The photon center-of-mass energy can be expressed in terms of $s$ and the
photon virtuality $k^2$ as
$$ k_0 = { s - m^2 + k^2 \over 2 \sqrt{s}}\,,\qquad\quad k_0^{\rm thr
} ={2m\over 1+ 2 \mu} \biggl[ \mu + \mu^2 + {\nu \over 4} \biggr] \eqno(2.5) $$
introducing a second small parameter
$$ \nu = {k^2 \over m^2} \eqno(2.6)$$
since we will consider only small photon four-momenta squared $|k^2| << m^2$.
When discussing experimental quantities like total cross sections, it is more
common to consider the laboratory system in which $N(p_1)$ is at rest. The
photon lab energy denoted by $E_\gamma$ is then
$$ E_\gamma = {s - m^2 - k^2 \over 2m} \eqno(2.7)$$
and the threshold value for two pion-photoproduction is given by
$$E_\gamma^{\rm thr} = 2 M_\pi( 1 + \mu) \,.\eqno(2.8)$$
For real photons only, we will calculate two-pion production amplitudes above
threshold.
In that case the pertinent four-momenta in the center-of-mass frame read
$$k^\mu = (k_0,\vec k), \quad p_1^\mu = (E_1,-\vec k),\quad
p_2^\mu =(E_2 , - \vec q_1 - \vec q_2),\quad q_1^\mu = ( \omega_1 , \vec q_1),
\quad q_2^\mu =   (\omega_2, \vec q_2) \eqno(2.9)$$
together with
$$\eqalign{ & \omega_i^2  = q_i^2 + M_\pi^2,\qquad
 E_1^2 = m^2 + k_0^2 , \qquad E_2^2 = m^2 + q_1^2 + q_2^2 + 2 q_1 q_2 z,
\cr  & q_1 \cdot k = k_0 ( \omega_1 - x q_1), \qquad q_2 \cdot k = k_0(
\omega_2 - y q_2)\cr} \eqno(2.10)$$
where $x,y,z$ denote the corresponding cosines of the angles between the
three-vectors and $k_0 = |\vec k|= (s-m^2)/2 \sqrt{s},\,\, q_i = |\vec q_i|$
give their lengths.
{}From spherical geometry one can deduce the standard relation
$$ y = xz + \sqrt{(1-x^2)(1-z^2)} \cos \varphi \eqno(2.11)$$
with $\varphi$ the azimuthal angle between the planes spanned by $\vec q_1$ and
$\vec k$ as well as $\vec q_1$ and $\vec q_2$. Finally, real photons satisfy
$\epsilon\cdot k = 0$. In the following, we will work in the Coulomb gauge
$\epsilon_0 = 0$, so that we have in photoproduction in addition $\vec \epsilon
\cdot \vec k = 0$
\medskip
\noindent{\bf II.3. THRESHOLD MULTIPOLES}
\medskip
At threshold in the center-of-mass frame ($i.e.\,\,\vec q_1 = \vec q_2 = 0$),
the two-pion electroproduction current matrix element can be decomposed into
multipole amplitudes as follows if we work to first order in the
electromagnetic coupling $e$,
$$\eqalign{ T \cdot \epsilon  = \chi_f^\dagger \bigl\{ & i \vec \sigma\cdot (
\vec \epsilon \times \vec k) \bigl[ M_1 \delta^{ab} + M_2 \delta^{ab} \tau^3 +
M_3 (\delta^{a3} \tau^b + \delta^{b3} \tau^a) \bigr] \cr  & + \vec \epsilon
\cdot \vec k  \bigl[ N_1 \delta^{ab} + N_2 \delta^{ab} \tau^3 + N_3
(\delta^{a3} \tau^b + \delta^{b3} \tau^a) \bigr] \bigr\} \chi_i \cr }
\eqno(2.12)$$
with $\chi_{i,f}$ two-component Pauli-spinors and isospinors and we used the
gauge $\epsilon_0 = 0$. Clearly, for real photons only the $M_{1,2,3}$ can
contribute. For virtual photons, gauge invariance $T\cdot k = 0$ allows to
reconstruct $T_0$ as $T_0 = \vec T \cdot \vec k /k_0$. The multipole amplitudes
$M_{1,2,3}$ and $N_{1,2,3}$ encode the information about the structure of the
nucleon as probed in threshold two pion photo- and electroproduction. The
physical channels listed in eq.(2.2) give rise to the following linear
combination of $M_{1,2,3}$ (and $N_{1,2,3}$ for $k^2 <0$).
$$\eqalign{ \gamma p & \to \pi^+ \pi^- p: \,\,M_1 + M_2, \cr
\gamma p & \to \pi^+ \pi^0 n: \,\,\sqrt{2} M_3, \cr \gamma p & \to
\pi^0 \pi^0 p:\,\,M_1 + M_2 + 2 M_3, \cr \gamma n & \to \pi^+ \pi^- n:\,\,
M_1 - M_2,\cr  \gamma n & \to \pi^0 \pi^- p:\,\,\sqrt{2} M_3, \cr
\gamma n & \to \pi^0  \pi^0 n:\,\, M_1 - M_2 - 2 M_3 \cr} \eqno(2.13)$$
which shows the abovementioned equality for the second and fifth channel.
\medskip
\noindent{\bf II.4. TOTAL CROSS SECTIONS}
\medskip
Let us consider the two-pion photoproduction close to threshold. The invariant
matrix element squared averaged over nucleon spins and photon polarizations
takes the form
$$|{\cal M}_{fi}|^2 = \vec k^2 \,| \eta_1 M_1 + \eta_2 M_2 + \eta_3 M_3|^2
\eqno(2.14)$$
with the isospin factors $\eta_{1,2,3}$ given in eq.(2.13).
The main dynamical assumption in this relation is that the two-pion
photoproduction amplitude in the threshold region can be approximated by the
amplitude at threshold.
Expressing $\vec
k^2$ in terms of $s$ and supplementing $|{\cal M}_{fi}|^2$ by the photon
flux factor $m^2/ p_1\cdot k= 2m^2 /(s-m^2)$, we find for the unpolarized total
cross section
$$\sigma_{\rm tot}^{\gamma N \to \pi\pi N}(s) = {m^2 \over 2s} (s-m^2)
\Gamma_3(s)\, |\eta_1 M_1 + \eta_2 M_2 + \eta_3 M_3|^2 \, S\,. \eqno(2.15)$$
Here, $\Gamma_3(s) $ is the integrated three-body phase space and $S$ a Bose
symmetry factor, $S = 1/2$ for the $\pi^0 \pi^0$ final state and $S = 1$
otherwise. The integrated three-body phase space can be expressed as
$$\eqalign{\Gamma_3(s) = & {1\over 32 \pi^3 } \int_0^{T_1}dT{\sqrt{T(T+2m)(T_1
- T)(T_2 -  T)} \over T_3 - T}\,,\cr T_1 = & {1\over 2 \sqrt{s}} ( \sqrt{s} -
m -  M_{\pi 1} - M_{\pi2} ) ( \sqrt{s} - m + M_{\pi 1} + M_{\pi 2})\,,\cr T_2 =
& {1\over 2 \sqrt{s}} ( \sqrt{s} -  m -  M_{\pi 1} + M_{\pi2} ) ( \sqrt{s} - m
 + M_{\pi 1} - M_{\pi 2} )\,,\cr T_3 = & {1\over 2 \sqrt{s}} ( \sqrt{s} - m)^2
 \cr} \eqno(2.16)$$
where $M_{\pi 1}$ and $M_{\pi 2}$ stand for the masses of the final state pions
and one has the inequality $ 0 \leq T_1 \leq T_{2,3}$. For equal pion masses an
excellent approximation to eq.(2.16) is given by
$$\Gamma_3(s) \approx { M_\pi m^{5/2} \over 64 \pi^2 ( m + 2 M_\pi)^{7/2}} \,
[E_\gamma - 2 M_\pi( 1 + \mu) ] ^2\,. \eqno(2.17)$$
Of course, an analogous approximation can be derived for unequal pion masses.
Consequently, the unpolarized total cross section can be approximated within a
few percent by the handy formula
$$\sigma_{\rm tot}^{\gamma N \to \pi \pi N}(E_\gamma)
\approx { M_\pi^2 ( 1 + \mu)
\over 32 \pi^2 ( 1 + 2\mu)^{11/2}} |\eta_1 M_1 + \eta_2 M_2 + \eta_3 M_3 |^2
\,S \, (E_\gamma - E_\gamma^{\rm thr})^2\,. \eqno(2.18)$$
For electroproduction, the prefactor in eq.(2.18) has
to be modified slightly to account for the virtual photon flux normalization
and then it gives the transverse total electroproduction cross section. For the
reaction $\gamma p \to \pi^+ \pi^- p$, we will also discuss results where  the
amplitude away from threshold is taken into account. In that case the total
cross section has to be worked out numerically. A most efficient integration
over the final state three--body phase space has the form
$$\sigma_{\rm tot}^{\gamma p \to \pi^+ \pi^- p} (E_\gamma) = {m \over 64 \pi^4
E_\gamma} \int\hskip - 5pt\int_{z^2<1} d \omega_1 d \omega_2
\int_{-1}^{+1} dx\int_0^\pi d \varphi |{\cal M}_{fi}|^2(k_0, \omega_1,\omega_2,
q_1,q_2,x,y,z) \eqno(2.19)$$
where we have chosen as the four independent variables which characterize the
final state configuration the pion energies $\omega_1$ and $\omega_2$, $x$ the
cosine of the angle between $\vec q_1$ and $\vec k$ and the azimuthal angle
$\varphi$ introduced in eq.(2.11). The cosine of the angle between $\vec q_1$
and $\vec q_2$ is already fixed by energy conservation to the value
$$z = z(\omega_1,\omega_2,s) = {\omega_1 \omega_2 - \sqrt{s} (\omega_1 +
\omega_2) + M_\pi^2 + {1\over 2}(s-m^2) \over \sqrt{(\omega_1^2 -
M_\pi^2)(\omega^2_2 - M_\pi^2)}} \,.\eqno(2.20)$$
Evidently, the quantity $z$ has to lie between $-1$ and $+1$ and this condition
determines the allowed region in the $\omega_1 \omega_2$ energy plane.  We have
also considered the first above threshold correction for $\gamma p \to \pi^0
\pi^0 p$ as detailed in appendix A.
This completes the necessary formalism.
\medskip
\noindent{\bf II.5. REMARKS ON ISOSPIN-BREAKING}
\medskip
In the calculations to be performed later on, it is legitimate to work with one
nucleon and one pion mass, which we will choose to be $m = m_p = 938.27$ MeV
and $M_\pi = M_{\pi^\pm} = 139.57$ MeV. However, the mass difference
$M_{\pi^\pm} - M_{\pi^0} = 4.6$ MeV in reality leaves a 11.9 MeV gap between
the production threshold of two neutral versus two charged pions. While we are
not in position of performing a calculation including all possible
isospin-breaking effects, a minimal procedure  to account for the mass
difference of the physical particles is to put in these by hand in the
pertinent kinematics, such that the thresholds open indeed at the correct
energy value. To be specific, for the $\pi^+ \pi^- p$ final state the threshold
photon energy is
$$E_\gamma^{\rm thr}(\pi^+ \pi^- p) = 320.66 \,{\rm MeV} \eqno(2.21)$$
whereas for $\pi^0 \pi^0 p$ it is
$$E_\gamma^{\rm thr}(\pi^0 \pi^0 p) = 308.77 \,{\rm MeV}\,. \eqno(2.22)$$
Also, in the pertinent three--body phase space integrals we will differentiate
between neutral and charged pion mass when we present results incorporating
the correct opening of the thresholds.
\goodbreak \bigskip
\noindent{\bf III. EFFECTIVE LAGRANGIAN}
\medskip
In this section, we will briefly discuss the chiral effective Lagrangian
underlying our calculation. We will also present an extension to incorporate
the $\Delta(1232)$ resonance. This is mandated by the closeness of the two-pion
production threshold and the location of the $\Delta$-resonance, $m_\Delta - m
- 2 M_{\pi^+} $ = 14.6 MeV. Concerning the chiral interactions of the $\pi N$
  system our discussion will be brief. Many additional details are spelled out
  in refs.[6,12].
\medskip
\noindent{\bf III.1. CHIRAL PION-NUCLEON LAGRANGIAN}
\medskip
To explore in a systematic fashion the consequences of spontaneous and explicit
chiral symmetry breaking of QCD, we make use of
 baryon chiral perturbation theory
(in the heavy mass formulation) [9] (HBCHPT). The nucleons are considered as
extremely heavy. This allows to decompose the nucleon Dirac spinor into "large"
$(H)$ and "small" ($h)$ components
$$\Psi(x) = e^{-i m v \cdot x } \{ H(x) + h(x)\} \eqno(3.1)$$
with $v_\mu$ the nucleon four-velocity, $v^2 = 1$, and the velocity eigenfields
are defined via $ \barre v H = H$ and $\barre v h = - h$.
Eliminating the "small" component field $h$ (which generates $1/m$
corrections), the leading order chiral $\pi N$ Lagrangian reads
$${\cal L}_{\pi N}^{(1)} = {\bar H} ( i v\cdot D + g_A S \cdot u ) H
\eqno(3.2)$$
Here the pions are collected in a SU(2) matrix-valued field $U(x)$
$$ U(x) = {1\over F} \bigl[ \sqrt{ F^2 - {\svec \pi}(x)^2 } + i \svec \tau
\cdot \svec \pi(x) \bigr] \eqno(3.3)$$
with $F$ the pion decay constant in the chiral limit and the so-called
$\sigma$-model gauge has been chosen which is of particular convenience for our
calculations in the nucleon sector. In eq.(3.2) $D_\mu = \partial_\mu +
\Gamma_\mu$ denotes the nucleon chiral covariant derivative, $S_\mu$ is a
covariant generalization of the Pauli spin vector, $g_A \simeq 1.26$ the
nucleon axial vector coupling constant (formally the one in the chiral limit)
and
$$u_\mu = i u^\dagger \nabla_\mu U u^\dagger \eqno(3.4)$$
with $u = \sqrt{U}$ and $\nabla_\mu$ the covariant derivative acting on the
pion fields. To leading order, ${\cal O}(q)$ one has to calculate tree
diagrams from
$$ {\cal L}^{(1)}_{\pi N} + {F^2 \over 4}{\rm Tr}\bigl\{ \nabla^\mu U
\nabla_\mu U^\dagger + \chi_+ \bigr\} \eqno(3.5)$$
where the second term is the lowest order mesonic chiral effective Lagrangian,
the non-linear $\sigma$-model coupled to external sources.
At next-to-leading order ${\cal O}(q^2)$ one has to consider tree graphs from
$$\eqalign{
{\cal L}_{\pi N}^{(2)} & =  {\bar H} \biggl\lbrace  -{1\over 2m} D \cdot D +
{1 \over 2m} (v \cdot D)^2 + c_1\Tr \chi_+  + \bigl( c_2 - {g_A^2
\over 8 m} \bigr)  v\cdot u\, v\cdot u + c_3\,   u\cdot u \cr & + c_4
[S^\mu,S^\nu] u_\mu u_\nu - {i g_A \over 2m} \lbrace S \cdot D , v \cdot u
\rbrace - { i\over 4 m}[S^\mu , S^\nu ] \bigl( (1 + \krig\kappa_v) f^+_{\mu
\nu} + {\krig\kappa_s - \krig\kappa_v  \over 2 } \Tr f^+_{\mu \nu} \bigr)
\biggr\rbrace H \cr} \eqno(3.6)$$
Some of the terms in eq.(3.6) are the $1/m$ corrections from the original Dirac
Lagrangian. But there are new terms proportional to the isoscalar and isovector
anomalous magnetic moments $\krig \kappa_{s,v}$ in the chiral limit or the low
energy constants $c_1,c_2,c_3,c_4$. The latter are related to the $\pi N$
$\sigma$-term and $\pi N$ scattering lengths.
In order to restore unitarity in a perturbative fashion, one has to include
(pion) loop diagrams. In HBCHPT, there exists a strict one-to-one
correspondence between the expansion of any observable in small external
momenta {\it and} quark masses and the expansion in the number of loops. In
what follows we will work within the one-loop approximation corresponding to
chiral power ${\cal O}(q^3)$.\footnote{*}{For more accurate calculations, it
seems to be necessary to include all terms of order $q^4$ as demonstrated for a
specific example in ref.[13].} To obtain all contributions at order $q^3$ one
has to supplement the chiral effective Lagrangian by additional terms ${\cal
L}^{(4)}_{\pi \pi} + {\cal L}^{(3)}_{\pi N} $ which will also serve to cancel
the divergences of certain loop diagrams. In our case of two-pion
electroproduction at threshold, however, all individual contributions at order
$q^3$ are finite.

The corrections at order $q^3$ to the two-pion threshold
photo/electroproduction amplitudes can be
grouped into three classes, $(i)$ the one-loop diagrams with
insertion from the leading order Lagrangian eq.(3.5), $(ii)$ kinematical
corrections with inverse powers of $m$ arising from the (chiral) expansion of
the relativistic nucleon pole graphs and $(iii)$ contact graphs from ${\cal
L}^{(3)}_{\pi N}$ with a priori unknown coefficients related to dynamics at
higher mass scales and tree graphs with one vertex from ${\cal L}^{(4)}_{\pi
\pi}$ ($e.g.$ the Wess-Zumino term which takes care about the anomalous Ward
identities of QCD). In estimating the a priori unknown coefficients of the
contact graphs we will make use of the resonance saturation principle and
relate them to the $\Delta(1232)$ exchange contributions. This is motivated by
the closeness of the $\Delta(1232)$ resonance and its relatively large
couplings to the $\gamma \pi N$ system. The effects of heavier baryon
resonances are neglected and vector meson exchange is suppressed by higher
powers of $q$ at threshold. Therefore we have now to give  the chiral $\pi N
\Delta$ Lagrangian to lowest order,
$${\cal L}^{(1)}_{\pi N\Delta} = {3 g_A \over 2 \sqrt{2}} \bar \Delta_\mu^a
\bigl[ g^{\mu\nu} - (Z + {1\over 2} ) \gamma^\mu \gamma^\nu\bigr] u^a_\nu \Psi
\quad + {\rm h.c.} \eqno(3.7)$$
with $u_\nu^a = {1\over 2} {\rm Tr}(\tau^a \, u_\nu)$ and we used the well
known SU(4) relation $g_{\pi N \Delta} = 3 g_{\pi N}/\sqrt{2} $ together with
the Goldberger-Treiman relation. The relativistic formulation using a
Rarita-Schwinger spinor for the $\Delta$ introduces  an off-shell parameter $Z$
into the $\pi N \Delta$ and $\gamma \pi N \Delta$ vertices which would be lost
in the nonrelativistic isobar formulation. It allows us to keep track of the
uncertainties in the $\Delta$ resonance contribution to certain observables.
Furthermore one has to consider the $\gamma N \to \Delta$ transition vertex
which is proportional to the electromagnetic field strength tensor. Since
the couplings of this vertex do not enter our
final result for the two-pion production  amplitudes at threshold calculated up
to order $q^3$, we do not explicitely give it here. The interested reader can
find a discussion on this topic in ref.[14].
We have now assembled all tools to perform a complete calculation up to and
including ${\cal O}(q^3) $ of two-pion photo/electroproduction.
\goodbreak \bigskip
\noindent{\bf IV. CHIRAL EXPANSION OF THE THRESHOLD AMPLITUDES}
\medskip
In this section, we will be concerned with the chiral expansion of the
threshold amplitudes $M_{1,2,3}$ and $N_{1,2,3}$. In each case we will give two
complete chiral powers, the leading and next-to-leading term. Before giving
the explicit expressions, a few general remarks are in order. Directly at
threshold the calculation simplifies enormously. In the center-of-mass frame
the final state nucleon and the two pions are at rest, $i.e.\,\, \vec q_1 =
\vec q_2 = 0$. With $v^\mu = (1,0,0,0)$ the spin-operator is the usual one
$S^\mu = ( 0, \vec \sigma/2 )$ and the two pion momenta are equal, $q^\mu_1 =
q^\mu_2 = (M_\pi,0)$. We therefore have in the Coulomb gauge $\epsilon_0 = 0$,
$$\epsilon \cdot v = \epsilon\cdot q_i = 0, \qquad S\cdot q_i = 0,\qquad v\cdot
(q_1 - q_2) = 0\,. \eqno(4.1)$$
In photoproduction, we have in addition $\epsilon \cdot k = 0$. Here, the
advantage of the heavy mass formulation (HBCHPT) clearly shows since the
conditions eq.(4.1) make most diagrams vanishing at threshold and only very few
diagrams are left to contribute. In words eq.(4.1) says, that the photon
nucleon vertex, the (out-going) pion nucleon vertex and the Weinberg vertex for
the two out-going pions all vanish at threshold. In photoproduction even the
photon coupling to an out-going pion becomes zero.  Furthermore, the $\gamma
\pi\pi NN $ contact term in ${\cal L}^{(1)}_{\pi N} $ is vanishing since it is
proportional to $\epsilon \cdot v$. Such selection rules at threshold are
extremely useful to fish out the non-vanishing one-loop graphs from a huge
number of diagrams. See also ref.[12] where such methods were applied
to nucleon Compton scattering and $\pi^0$ photoproduction at threshold
demonstrating their  power.
\medskip
\noindent{\bf IV.1. THE TRANSVERSE THRESHOLD AMPLITUDES}
\medskip
Let us first discuss the transverse amplitudes $M_{1,2,3}$. In this case we can
still exploit the condition $\epsilon \cdot k = 0$ but we will not restrict
$k^2$ to
be zero in order to get also the result for electroproduction. The various
contributions to $M_{1,2.3}$ are ordered by their chiral power. To leading
order ${\cal O}(q)$ we have the tree graphs from ${\cal L}^{(1)}_{\pi N}$ and
one finds quite easily that due to the abovementioned selection rules all these
graphs are exactly zero at threshold. The leading nonzero contributions
therefore comes from tree graphs with one insertion from ${\cal L}^{(2)}_{\pi
N}$. The two non-vanishing diagrams are shown in fig.2. It is interesting to
note that the first non-vanishing term for two-pion photoproduction at
threshold start at order $q^2$ in all channels. This is very different from
single pion photoproduction where the charged channels are strongly enhanced.

Next, we have to study all contributions at chiral power ${\cal O}(q^3)$. Among
these are the one-loop contributions. Again the selection rules eq.(4.1) and
$\epsilon\cdot k = 0$ reduce drastically the number of  graphs which are
nonzero at threshold  to four. These are depicted in fig.3.  Finally we have to
work out all polynomial contributions at order $q^3$. Part of these (the
$1/m^2$ corrections) can be most easily obtained if we calculate the nucleon
pole graphs relativistically and then expand in powers of $M_\pi$ and $k^2$.
Further terms are given by tree graphs involving two vertices from ${\cal
L}^{(2)}_{\pi N}$ where at least one carries a constant $c_1, c_2 , c_3, c_4$
or $\krig \kappa_s, \krig \kappa_v$. The two posssible graphs are shown in
fig.4a, their sum, however, vanishes since the nucleon propagators in both have
opposite sign. Then we have to study tree graphs with one vertex from ${\cal
L}^{(4)}_{\pi\pi}$. Most of  these insertions just lead to a multiplicative
renormalization of the ${\cal O}(q)$ graphs, which are, however, all equal to
zero. ${\cal L}^{(4)}_{\pi \pi} $ contains the Wess-Zumino term incorporating
the anomalous (natural parity violating) vertex $\gamma \to 3 \pi$.
Nevertheless, the diagram shown in fig.4b vanishes at threshold, simply
because it is proportional to $\epsilon_{\mu\nu\alpha\beta}q_1^\alpha \,
q_2^\beta = 0$ with $q_1 = q_2$. Finally, in order to complete
the list of all possible contributions at ${\cal O}(q^3)$ we have
to consider the
(genuine) contact terms from ${\cal L}^{(3)}_{\pi N}$. These carry new low
energy constants which are not known a priori and have to be determined from
phenomenology. We will invoke here the resonance saturation principle (which
has been tested in the meson sector to work very well) and estimate these
constant from the $\Delta(1232)$-resonance contribution to two-pion
photo/electroproduction at threshold. One expects sizeable effects from the
$\Delta(1232)$ since first it is quite close to threshold and second its
couplings to the $\gamma \pi N$ system are very large (about twice the nucleon
couplings). On first sight the distance of only $14.6 $ MeV of the
$\Delta(1232)$ from threshold seems to give rise to overwhelming contributions
since one naively expects that the very small denominator
$${1\over m_\Delta^2- s_{\rm thr}}
= {1 \over m_\Delta - m - 2 M_\pi}\,\, {1 \over
m_\Delta + m + 2 M_\pi} \eqno(4.2)$$
enters the result. We have worked out all possible graphs with a single or
double $\Delta$ excitation (cf. fig.5) and found that at threshold the small
and dangerous
denominator $m_\Delta - m - 2 M_\pi$ always gets cancelled by  exactly the
same term in the numerator.  This is a very important feature since otherwise
the $M_\pi$ expansion would be in serious trouble. $M_\pi/(m_\Delta - m - 2
M_\pi) $ formally counts as small, ${\cal O}(q)$, but numerically it is of
course very large. After this detailed discussion of the chiral expansion of
the transverse threshold amplitude, let us now give the final result,
$$\eqalign{
M_1 = &  {e g_A^2 M_\pi \over 4 m^2 F_\pi^2} + {\cal O}(q^2)\,, \cr
M_2 = & {e \over 4 m F_\pi^2} (2g_A^2 - 1 - \kappa_v) + {e M_\pi \over 4 m^2
F_\pi^2} (g_A^2 - \kappa_v) - {e g_A^2 M_\pi \over 8 m m_\Delta^2 F_\pi^2}
B_\Delta \cr & + {e g_A^2 M_\pi \over 64 \pi F_\pi^4} \biggl\{ {8 + 4r \over
\sqrt{1+r}} \arctan\sqrt{1+r} - {r \over 1 + r} -{ 1+ r+r^2 \over
(1+r)^{3/2}}\biggl[ {\pi \over 2 } + \arctan{r \over \sqrt{1 + r}} \biggl]
\cr & + i \biggl[ {\sqrt{3}(2+r) \over 1 + r} - {1 + r + r^2 \over (1+r)^{3/2}}
\ln { 2 + r + \sqrt{3(1+r)} \over \sqrt{1 + r +r^2}} \biggr] \biggr\} + {\cal
O}(q^2)\,, \cr
M_3 = & {e \over 8 m F_\pi^2} (1 + \kappa_v- 2 g_A^2) + {e M_\pi \kappa_v\over
8 m^2 F_\pi^2} + {e g_A^2 M_\pi \over 16 m m_\Delta^2 F_\pi^2}
B_\Delta \cr & + {e g_A^2 M_\pi \over 256 \pi F_\pi^4} \biggl\{6- {4 + 2r \over
\sqrt{1+r}} \arctan\sqrt{1+r} - {r \over 1 + r} -{ 1+ r+r^2 \over
(1+r)^{3/2}}\biggl[ {\pi \over 2 } + \arctan{r \over \sqrt{1 + r}} \biggl]
\cr & + i \biggl[ {\sqrt{3}(2+r) \over 1 + r} - {1 + r + r^2 \over (1+r)^{3/2}}
\ln { 2 + r + \sqrt{3(1+r)} \over \sqrt{1 + r +r^2}} \biggr] \biggr\} + {\cal
O}(q^2) \cr } \eqno(4.3)$$
with the ratio $r = -k^2/4M_\pi^2$
and
$$B_\Delta = {2m_\Delta^2 + m_\Delta m - m^2 \over m_\Delta - m} + 4Z [
m_\Delta(1+2Z) + m(1+Z)] \eqno(4.4)$$
which involves the off-shell parameter $Z$ of the $\pi N \Delta$ vertex. In
fact, taking the allowed range of $Z$ given in ref.[14], $i.e.\,\, -0.8 < Z <
0.3$ we find a weak $Z$-dependence, $i.e.$ 9.9 GeV $< B_\Delta < 15.1$ GeV.
Furthermore, from the isospin factors of eq.(2.13) we see that to order $M_\pi$
the $\Delta$ contributions are absent in the $\pi^0 \pi^0$ channels.

It is interesting to note that $M_1$ is zero to ${\cal O}(q^0)$ and
${\cal O}(q)$. At next order
$q^2$ it only receives a contribution from the relativistic $N$-pole graphs,
but no loop or counter term contribution. The result eq.(4.3) has been written
completely in terms of  physical parameters, not the chiral limit values which
enter the effective Lagrangian. For $F_\pi, M_\pi, g_A$ and $ m$
 these differences
only would show up at the next order ${\cal O}(q^4)$. The isovector anomalous
magnetic  moment, however, has been renormalized and includes a non-analytic
piece $\sim \sqrt{\hat m}$ [12],
$$\kappa_v = \krig \kappa_v - { g_A^2 m M_\pi\over 4\pi F_\pi^2}\,.\eqno(4.5)$$

Note that the first terms in the expansion of $M_{2,3}$ arising at order $q^2$
differ from the corresponding expressions given in ref.[10] by the term
proportional to $\kappa_v$. In that reference, the Weinberg Lagrangian was used
and the photon was coupled in via minimal substitution. Such a procedure can,
however, not generate the photon nucleon coupling proportional to the anomalous
magnetic moment. It is furthermore
important to write down such an anomalous photon nucleon vertex in a manifestly
chiral invariant fashion using in eq.(3.6) the quantity $f^+_{\mu\nu}$ and not
just the photon field strength tensor. When expanding $f^+_{\mu\nu}$ in powers
of the pion field the pertinent $\gamma \pi \pi NN$ vertex (proportional to
$\krig \kappa_v$) based on chiral symmetry is automatically generated.

Another point worth mentioning is that the transverse multipoles
$M_{2,3}$ are $k^2$-dependent only through their loop contribution. This can
be understood from the fact that the tree graphs have to be polynomial in both
$M_\pi$ and $k^2$ and that a term linear in $k^2$ is already of higher order in
the chiral expansion. It is also not possible to further expand the
$r$-dependent functions since $r = -k^2/4 M_\pi^2$ counts as order one and all
terms have to be kept.

At first sight the multipoles $M_{2,3}$ seem to behave singular in the chiral
limit $M_\pi = 0$ since then $r$ becomes infinite. This is, however, not the
case and the chiral limit is perfectly smooth (as it should be) with
$$\krig M_2^{loop} = - 2\krig M_3^{loop} = {e \krig g_A^2 \over 128 F^4}
\sqrt{-k^2} = {e \over 4 \krig m F^2} \bigl[ \krig \kappa_v - \krig F_2^v(k^2)
\bigr] \eqno(4.6)$$
where $\krig F_2^v(k^2)$ is the nucleon isovector magnetic form factor in the
chiral limit calculated up to order $q^3$ (see ref.[12]).

The loop contribution to the transverse  multipoles of two pion
production as given in eq.(4.3) have a nonzero imaginary part even at
threshold. This comes from the rescattering type graphs. Due
to unitarity the pertinent loop functions have a right hand cut starting at $s
= (m+ M_\pi)^2$ (the single pion production threshold) and these functions are
here  to be evaluated at $s = (m+2M_\pi)^2$ (the two-pion production
threshold). For photoproduction the complicated $k^2$-dependence of the loop
terms disappears and we have the complex constants
$$\eqalign{M_2^{loop} & = {eg_A^2 M_\pi\over 64 \pi F_\pi^4}\biggl\{{3\pi \over
2} + i \bigl[ 2 \sqrt{3} - \ln(2+\sqrt{3})\bigr] \biggr\} \cr
M_3^{loop} & = {e g_A^2 M_\pi \over 256 \pi F_\pi^4} \biggl\{ 6 - {3\pi \over
2} + i \bigl[ 2 \sqrt{3} - \ln(2+\sqrt{3})\bigr] \biggr\}\,. \cr } \eqno(4.7)$$
This completes the discussion of the transverse threshold multipoles
up-to-and-including order $q^3$. \vfill \eject
\medskip
\noindent{ \bf IV.3. LONGITUDINAL MULTIPOLES AT THRESHOLD}
\medskip
In the electroproduction case, we also have the longitudinal threshold
amplitudes $N_{1,2,3}$. Since we can no more exploit the condition $\epsilon
\cdot k = 0$ the photon coupling to an out-going pion line is non-vanishing and
therefore we obtain a nonzero contribution already at leading order ${\cal
O}(q)$ involving a pion propagator (cf. fig.6a).
 Adding up all terms which arise at order
$q$ and $q^2$ we find the following results
$$\eqalign{N_1 & = {\cal O}(q)\,, \cr N_2 & = {e M_\pi( 1 + \mu)
\over F_\pi^2 ( 4 M_\pi^2 - k^2)} + { e(2g_A^2 - 1) \over 4 m F_\pi^2} + {\cal
O}(q)\,, \cr N_3 & = - {1\over 2} N_2 + {\cal O}(q)\,. \cr} \eqno(4.8)$$
It is interesting to note that none of the low energy constants
$c_1,c_2,c_3,c_4$ and the anomalous magnetic moments which enter ${\cal
L}^{(2)}_{\pi N}$ show up in the final result. The graph of fig.6b contains
the higher derivative $\pi \pi NN$ vertex proportional to $c_1,c_2,c_3$ but
the isospin factor $\delta^{ac} \epsilon^{c3b} + (a \leftrightarrow b) =0$
annihilates this diagram. With eq.(4.8) we have given two full chiral powers
(leading and next-to-leading) for the expansion of the longitudinal threshold
amplitudes $N_{1,2,3}$. We do not work out here the third chiral power,
which comes from loops and polynomial counter terms. We will now return to
photoproduction and present our results for observables like total cross
sections.
\goodbreak \bigskip
\noindent{\bf V. RESULTS AND DISCUSSION}
\bigskip
In this section, we will first present numerical results for the threshold
multipoles and total cross sections in the isospin limit. For the experimental
implications, we also consider the kinematical shifts for the various channels
as discussed in section 2.5.
\bigskip
\noindent{\bf V.1. RESULTS IN THE ISOSPIN LIMIT}
\bigskip
First, we must fix parameters. We work with $m = m_p$, $M_\pi = M_{\pi^\pm}$,
$F_\pi = 93$ MeV, $g_{\pi N}^2 / 4 \pi = 14.28$ and $e^2 / 4 \pi = 1/ 137.036$.
In fig.7, we show the invariant matrix--elements $|{\cal M}_{ij}|$ for the
$\gamma p$ initial state versus the off--shell parameter $Z$ defined in
eq.(3.7) ($k^2 = 0$). As stressed after eq.(4.4), the $\Delta$ does not
contribute to the $\pi^0 \pi^0 $ final state. For the allowed range of $Z$
taken from ref.[14], the $Z$--dependence is weak. The surprising result is
the dominance of the $\pi^0 \pi^0$ final state for the threshold multipoles
chirally expanded to ${\cal O}(M_\pi)$. In table 1, we give the ${\cal O}(1)$
and ${\cal O}(1) + {\cal O}(M_\pi)$ contributions with and without the $\Delta$
term. To lowest order, the $\pi^0 \pi^0$ final states are suppressed, as
already noted in ref.[10]. It is interesting to compare these results to the
case of single pion photoproduction. In that case, the Kroll--Ruderman term
dominates charged production and the loop effects are small. For $\gamma p \to
\pi^0 p$, however, the loop corrections are quite large [3]. The difference to
the two pion production case is that the total cross sections for
photoproducing a charged pion are nevertheless much larger than the
corresponding ones for neutral pions (in the threshold region). Notice also
that for the $\pi^+ \pi^-$ final state
the ${\cal O}(M_\pi)$ loop and kinematical corrections are largely
cancelled by the ones from the $\Delta$. In case of electroproduction, the
$k^2$--dependence of the $M_i$ $(i=1,2,3)$ is weak. For $k^2$ ranging from
0 to --0.1 GeV$^2$, the $|{\cal M}_{ij}|$
(in GeV$^{-3}$) change from 11.2 to 9.4, 20.1 to
15.1 and 35.2 to 29.7 for $\gamma p \to \pi^+ \pi^- p$, $\pi^+ \pi^0 n$ and
$\pi^0 \pi^0 p$, respectively. The corresponding longitudinal multipoles
$N_i$ $(i=1,2,3)$ are completely dominated by the pion pole (compare eq.(4.8))
and are not shown here.

In fig.8, we show the total cross sections for the first three channels given
in (2.2) by approximating the amplitudes in the threshold region
 through their threshold values. For $E_\gamma = 330$ MeV, we
find\footnote{$^*$}{These numbers differ from the ones reported in ref.[15]
since in that reference the $\Delta$--contribution was omitted.}
$$\eqalign{
 \sigma_{\rm tot} (\gamma p \to \pi^0 \pi^0 p) &= 0.36 \, {\rm nb}, \cr
 \sigma_{\rm tot} (\gamma p \to \pi^+ \pi^0 n) &= 0.22 \, {\rm nb}, \cr
 \sigma_{\rm tot} (\gamma p \to \pi^+ \pi^- p) &= 0.08 \, {\rm nb}\,\, . \cr}
 \eqno(5.1)$$
These are very small cross sections, but we refer to section 5.2. for further
discussion. In fig.9, we compare the lowest order cross section for the
production of two charged pions with the first correction from ${\cal O}(q)$,
cf. appendix A. While the former goes like $(E_\gamma - E_\gamma^{\rm thr})^2$,
the latter approximatively rises as $(E_\gamma - E_\gamma^{\rm thr})^3$. At
$E_\gamma \simeq 323$ MeV, these two contributions are of the same size. This
narrows the window in which the $\pi^0 \pi^0$ final state is dominant to the
first few MeV above threshold. The first correction to this particular channel
is rather small as discussed below (cf. appendix A).
 Let us now, however, review these results in
the light of the different thresholds already mentioned a few times.
\bigskip
\noindent{\bf V.2. EXPERIMENTAL IMPLICATIONS}
\bigskip
To connect to the experimental situation, we now consider the three--body
phase space with the physical masses for the corresponding pions. This
automatically takes care of the various threshold energies. In the loops
we work, however, with one pion mass since that effect is small as discussed
in appendix B. Chiefly, the loop functions are sensitive to $(M_{\pi^+} -
M_{\pi^0}) / M_{\pi^+} = 0.03$, i.e. a few percent effect. Fig.10 shows the
calculations with the correct phase--space and
using the threshold matrix--elements.
For $\gamma p \to \pi^+ \pi^- p$, we also show the first correction above
threshold from ${\cal L}_{\pi N}^{(1)}$. At $E_\gamma = 320$ MeV, the total
cross section for $\pi^0 \pi^0$ production is 0.5 nb  whereas the competing
$\pi^0 \pi^+ n$ final state has $\sigma_{\rm tot} = 0.07$ nb.
 Double neutral pion
production reaches $\sigma_{\rm tot} = 1.0$  nb at $E_\gamma = 324.3$ MeV
in comparison to $\sigma_{\rm tot} (\gamma p \to \pi^0 \pi^+ n) = 0.26$ nb and
$\sigma_{\rm tot} (\gamma p \to \pi^+ \pi^- p) < 0.1$ nb.
 This means that for the
first 10...12 MeV above $\pi^0 \pi^0$ threshold, one has a fairly clean signal
and much more neutrals than expected. Remember that to leading order
in the chiral expansion, the production of two neutral pions is
extremely suppressed. Of course, the above threshold correction for this
channel, which comes from ${\cal L}_{\pi N}^{(2)}$ (and higher orders)
should be calculated systematically. The first correction, which vanishes
proportional to $|{\svec q}_i|$ ($i=1,2$) at threshold, has been calculated
(see appendix A) and found to be very small. The corresponding cross section
at $E_\gamma =320$, $325$ and $330$ MeV is $\sigma_{\rm tot}^{\rm first \,
corr}
= 0.009, \, 0.026$ and $0.056$ nb, i.e a few percent of the leading order
result.
It is, therefore, conceivable that the qualitative features
described above will not change if even higher order corrections are taken
into account.
\goodbreak \bigskip
\noindent{\bf VI. BRIEF SUMMARY AND OUTLOOK}
\bigskip
In this paper, we have performed an exploratory study of two--pion photo-- and
electroproduction off the nucleon. We have focused on the threshold region and
used heavy baryon chiral perturbation theory to calculate the first two powers
in the chiral expansion of the transversal ($M_{1,2,3}$)
 and longitudinal ($N_{1,2,3}$) mulitpoles defined in eq.(2.12). The pertinent
results (we focus on the photoproduction case) are:
\medskip
\item{(i)}To leading order in the chiral expansion, only the multipoles $M_2$
and $M_3$ are non--vanishing, with $M_2 = -2M_3$. Therefore, the production of
two neutral pions is strictly suppressed.
\smallskip
\item{(ii)}At next order in the chiral expansion, one has to consider
kinematical (1/m) corrections, one--loop contributions and tree graphs with
insertion of the $\Delta (1232)$ (chirally expanded). The loop contributions
make the $\pi^0 \pi^0$ final state dominant in contrast to the expectation
from single pion production.
\smallskip
\item{(iii)}We have calculated total cross sections for the various channels
eq.(2.2) under the assumption that the amplitudes in the threshold region can
be approximated by the exact threshold amplitude. Taking into account the
different pion masses in the respective phase spaces, these cross sections are
shown in fig.10. There is a window of about 10 MeV above $\pi^0 \pi^0$
threshold in which one can detect much more neutrals than expected.
\smallskip
\item{(iv)}We have considered the first above threshold corrections for the
processes $\gamma p \to \pi^+ \pi^- p$ and $\gamma p \to \pi^0 \pi^0 p$. While
the former are sizeable even very close to threshold, the latter are rather
small. We conclude that ultimately one should
consider the full amplitude above threshold (including also more dynamics
from the resonance region) to draw more quantitative
conclusions.  We hope to report on this in the not too far future.
\goodbreak \bigskip \bigskip
\noindent{\bf APPENDIX A: LEADING ORDER AMPLITUDE ABOVE THRESHOLD}
\medskip
In section IV, we considered the two-pion photoproduction amplitude exactly at
threshold and we have seen that the selection rules make the ${\cal O}(q)$
contribution exactly vanishing at threshold  in all channels. Here we will now
work out the amplitude for the process $\gamma p \to \pi^+ \pi^- p$ (with
$k^2=0)$ away from threshold to leading order ${\cal O}(q)$. For the $\pi^0
\pi^0$ channels to this order the amplitude above threshold is still equal to
zero everywhere . The five diagrams shown in fig.11 (plus their crossed
partners
with the pion lines interchanged) give rise to the following amplitude, with
the
kinematics  taken over from eq.(2.9),
$$\eqalign{ T\cdot \epsilon =  {e \over 2F_\pi^2} \biggl\{ & - \vec \epsilon
\cdot \vec q_1 { \omega_2 \over q_1 \cdot k}  - \vec \epsilon \cdot \vec q_2 {
\omega_1 \over q_2 \cdot k}  +{g_A^2 \over \omega_2} \vec \sigma \cdot \vec
q_2\,\,\vec \sigma \cdot \vec \epsilon +{g_A^2 \over \omega_1} \vec \sigma
\cdot \vec \epsilon \,\,\vec \sigma \cdot \vec q_1 \cr & + {g_A^2 \vec \epsilon
\cdot \vec q_1 \over \omega_2 q_1 \cdot k} \vec \sigma \cdot \vec q_2\,\, \vec
\sigma \cdot (\vec k - \vec q_1) + {g_A^2 \vec \epsilon \cdot \vec q_2 \over
\omega_1 q_2 \cdot k} \vec \sigma \cdot(\vec k - \vec q_2)  \,\vec \sigma \cdot
\vec q_1\biggr\}\cr } \eqno(A.1)$$
To calculate the unpolarized total cross section we have to average over
initial spins and sum over final spins, $i.e.$ we must compute the quantity
$$ {1\over 2} {\rm Tr}[ T\cdot \epsilon \,\,(T\cdot \epsilon)^\dagger]
\eqno(A.2)$$
where the trace is taken over  $2\times 2 $ spin-matrices. From (A.2) we get a
quadratic form in
the polarization vector $\vec \epsilon$ which we have to average
over the two photon polarizations perpendicular to $\vec k$. The pertinent
polarization averages are done via the substitutions
$$\eqalign{ & (\vec \epsilon \cdot \vec q_1)^2 \to {1\over 2} q_1^2( 1 - x^2),
\cr & (\vec \epsilon \cdot \vec q_2)^2 \to {1\over 2} q_2^2 ( 1 - y^2),
\cr & \vec \epsilon \cdot \vec q_1\,\vec \epsilon \cdot \vec q_2 \to {1\over 2}
q_1 q_2 (z - xy), \cr & (\vec \epsilon\,)^2 \to 1\,. \cr } \eqno(A.3)$$
with $x,y,z$ the cosines of the inclined angles. Therefore, the invariant
matrix element which enters the cross section integral eq.(2.19) takes the form
$$\eqalign{
&|{\cal M}_{fi}|^2 =  {e^2 \over 8 F_\pi^4} \biggl\{
{q_1^2 \omega_2^2 \over (q_1\cdot k)^2 } (1- x^2) + 2 {q_1 q_2 \omega_1
\omega_2\over q_1\cdot k\, q_2 \cdot k } (z- xy)  + {q_2^2 \omega_1^2 \over
(q_2\cdot k)^2 } (1- y^2)  \cr
 & - 2 g_A^2 \biggl[ q_1^2 (1-x^2) {\omega_2
\over q_1 \cdot k} \biggl({1\over \omega_1} + { q_2(k_0 y - q_1z)
\over\omega_2 q_1 \cdot k} \biggr) + q_2^2 (1-y^2) {\omega_1 \over q_2 \cdot k}
\biggl({1\over \omega_2} + { q_1(k_0 x - q_2 z) \over\omega_1 q_2 \cdot k}
\biggr) \cr  & +  q_1 q_2 (z-xy) \biggl( {\omega_2
\over q_1 \cdot k} \biggl({1\over \omega_2} + { q_1(k_0 x - q_2 z)
\over\omega_1 q_2 \cdot k} \biggr) + {\omega_1 \over q_2 \cdot k}
\biggl({1\over \omega_1} + { q_2(k_0 y - q_1 z) \over \omega_2 q_1 \cdot k}
\biggr)\biggr) \biggr] \cr & + g_A^4 \biggl[ {2q_1^2 \over \omega_1^2} +{2q_2^2
\over \omega_2^2} + {q_1^2 q_2^2(1-x^2) \over \omega_2^2(q_1\cdot k)^2} (k_0^2
- 2 k_0 \omega_1 +q_1^2) + {q_1^2 q_2^2(1-y^2) \over \omega_1^2(q_2\cdot k)^2}
(k_0^2 - 2 k_0 \omega_2 +q_2^2) \cr &  + {2q_1 q_2^2 \over \omega_1 \omega_2
q_2\cdot k} \bigl( k_0(x+yz-2xy^2)+q_2(xy-z) \bigr) +{2q_1^2 q_2 \over \omega_1
\omega_2 q_1\cdot k} \bigl( k_0(y+ xz -2x^2y) + q_1(xy-z)\bigr) \cr & - {4q_1
q_2 \over \omega_1 \omega_2}xy + {2q_1^2 q_2^2(z-xy) \over
\omega_1 \omega_2 q_1 \cdot k \, q_2\cdot k} \bigl( k_0^2(2xy-z) - k_0q_1y -
k_0q_2x  +q_1q_2 \bigr) \biggr] \biggr\} \cr } \eqno(A.4)$$
The lengthy expression of the invariant matrix element from the rather simple
amplitude (A.1) makes clear that calculations above threshold are rather
voluminous. In fact, for the first 10 MeV above threshold the last two terms
in eq.(A.1) could be safely neglected since they vanish with a higher power
of $|\svec{q}_i|$ $(i=1,2)$.
In order to represent the most general amplitude above threshold
more than a dozen of amplitude functions are needed to characterize a specific
channel (see ref.[10]) and each of them depends on five independent variables
($e.g.$ $E_\gamma, \omega_1, \omega_2, x,y$). In comparison to this enormous
complexity single pion photoproduction is rather simple. The process is
completely described in terms of four amplitude functions which depend on
$E_\gamma$ and $e.g.$ the cms scattering angle.

Let us briefly turn to the first above threshold correction for $\gamma p \to
\pi^0 \pi^0 p$. The corresponding tree diagrams must include at least one
insertion from ${\cal L}_{\pi N}^{(2)}$ as noted before. From these, the ones
with the $\gamma \pi NN$ vertex from ${\cal L}_{\pi N}^{(2)}$ go to zero as
$|{\svec q}_i|$ $(i=1,2)$ as one approaches threshold whereas the ones with
a photon--nucleon vertex from ${\cal L}_{\pi N}^{(2)}$ and two pion--nucleon
vertices from ${\cal L}_{\pi N}^{(1)}$ go as $|{\svec q}_1| \, |{\svec q}_2|$,
i.e. they vanish faster. So to leading order in small momenta, we consider only
the first type of diagrams and find
$$ T \cdot \epsilon = { i e  g_A^2 \over 2 m F_\pi^2} \biggl[
{\omega_1 \over \omega_2} {\svec \sigma} \cdot ({\svec \epsilon} \times {\svec
q}_2 ) +
{\omega_2 \over \omega_1} {\svec \sigma} \cdot ({\svec \epsilon} \times {\svec
q}_1 ) \biggr] \eqno(A.5)$$
and from this the total cross section follows as
$$\sigma^{\gamma p \to \pi^0 \pi^0 p}_{\rm tot} (E_\gamma )
= {e^2 g_A^4 \over 384 \pi^3
E_\gamma m F_\pi^4} \int \int_{z^2 < 1} d\omega_1 \, d\omega_2 \, \biggl(
{\omega_1^2 q_2^2 \over \omega_2^2} +
{\omega_2^2 q_1^2 \over \omega_1^2} +  2 q_1 q_2 z \, \biggr) \quad .
 \eqno(A.6)$$
In this case, one can perform the $x$ and $\varphi$ integrations
in $|{\cal M}_{fi}|^2$ analytically.
We have, however, also performed the four--dimensional integration numerically
as a check and found good agreement with the results based on (A.6).
\goodbreak \bigskip
\noindent{\bf APPENDIX B: LOOP FUNCTIONS}
\medskip
The calculation of the one-loop graphs shown in fig.3 leads to the following
loop integrals (in $d$ space time dimensions):
$$\eqalign{ {1\over i} & \int {d^d l \over (2\pi)^d} { l_\mu l_\nu \over v\cdot
l (M_\pi^2 - l^2) (M_\pi^2 - (l+k)^2) }   = g_{\mu \nu} \gamma_3(\omega,k^2) +
\dots, \cr  {1\over i} & \int {d^d l \over (2\pi)^d} { l_\mu l_\nu \over v\cdot
l (M_\pi^2 - l^2) (M_\pi^2 - (l+k)^2)(M_\pi^2 - (l+k-2q)^2) }  = g_{\mu \nu}
Q_3(\omega,M_\pi,k^2) + \dots \cr} \eqno(B.1)$$
with $\omega = v\cdot k$ and $q = M_\pi\, v$. All propagators are understood to
have an additional infinitesimal negative imaginary part in the denominator,
which makes the Wick-rotation to euclidean space time unique. The ellipsis
stand
for terms proportional to $v_\mu v_\nu,\, k_\mu k_\nu$ etc., which are not
needed. Using standard Feynman parameter representations we obtain for the
symmetric sums at $d=4$ which enter the threshold amplitudes $M_{2,3}$
$$ \eqalign{ & \gamma_3(\omega,k^2) + \gamma_3(-\omega,k^2) = {1 \over 8 \pi}
\int_0^1 dx \sqrt{M_\pi^2 - x^2 \omega^2 + x(x-1) k^2}\,, \cr
& Q_3(\omega,M_\pi,k^2) + Q_3(-\omega,-M_\pi, k^2) = {1 \over 32 \pi
M_\pi(\omega- M_\pi)} \cr  & \times
\int_0^1 dx \bigl\{\sqrt{M_\pi^2 - x^2 \omega^2 + x(x-1) k^2}- \sqrt{M_\pi^2 +
4x M_\pi(\omega - M_\pi) - x^2 \omega^2 + x(x-1)k^2} \bigr\}. \cr} \eqno(B.2)$$
These are then to be evaluated at $ \omega = 2M_\pi + i0$. Let us finally
comment on isospin breaking effects.  If we consider $\pi^0\pi^0 $ production
the pions in the loop are all charged, therefore $M_\pi = M_{\pi^+}$ in
eq.(B.2) and the functions are to be evaluated at $\omega = 2 M_{\pi^0} + i0$.
Since both real and imaginary part of the functions in eq.(B.2) are smooth
around $\omega = 2M_{\pi^+}$ with finite first derivative the isospin breaking
effect from the different pion masses is of order $(M_{\pi^+} - M_{\pi^0})/
M_{\pi^+} \simeq 3\%$ and therefore very small. This is quite different to the
reaction $\gamma p \to \pi^0 p$ where the
respective loop functions vary strongly around the branch point $\omega =
M_{\pi^+}$ (of square root type) and the effective isospin breaking parameter
is $\sqrt{(M_{\pi^+} - M_{\pi^0}) / M_{\pi^+}} \simeq 18\%$.
\goodbreak \bigskip \bigskip \vfill \eject
\noindent{\bf REFERENCES} \bigskip
\item{1.}E. Mazzucato {\it et al.},
 {\it Phys. Rev. Lett.\/} {\bf 57} (1986) 3144;

R. Beck {\it et al.}, {\it Phys. Rev. Lett.\/} {\bf 65} (1990) 1841.
\smallskip
\item{2.}T. P. Welch et al., {\it Phys. Rev. Lett.\/}
{\bf 69} (1992) 2761.
\smallskip
\item{3.}V. Bernard, N. Kaiser, and Ulf-G. Mei{\ss}ner, {\it Nucl. Phys.\/}
{\bf B383} (1992) 442; {\it Phys. Rev. Lett.\/} {\bf 67} (1991) 1515;
$\pi N$ {\it Newsletter} {\bf 7} (1992) 62;
{\it Phys. Lett.\/} {\bf B282} (1992) 448;

V. Bernard, N. Kaiser, T.--S. H. Lee and Ulf-G. Mei{\ss}ner,
{\it Phys. Rev. Lett.\/} {\bf 70}

 (1993) 367; preprint BUTP--93/23, 1993, to app. in {\it Phys. Rep.}.
\smallskip
\item{4.}J.M. Laget, {\it Phys. Rep.\/} {\bf 69} (1981) 1.
\smallskip
\item{5.}E. Amaldi, S. Fubini and G. Furlan,
{\it Pion--Electroproduction},
Springer Verlag, Berlin 1979.
\smallskip
\item{6.}Ulf-G. Mei{\ss}ner,
{\it Rep. Prog. Phys.\/} {\bf 56} (1993) 903.
\smallskip
\item{7.}T.--S. Park, D.--P. Min and M. Rho,
 {\it Phys. Rep.\/} {\bf 233} (1993) 341.
\smallskip
\item{8.}S. Weinberg, {\it Physica} {\bf 96A} (1979) 327. \smallskip
\item{9.}E. Jenkins and A.V. Manohar, {\it Phys. Lett.\/} {\bf B255} (1991)
558.
\smallskip
\item{10.}R. Dahm and D. Drechsel, in Proc. Seventh Amsterdam Mini--Conference,
eds. H.P. Blok, J.H. Koch and H. De Vries, Amsterdam, 1991. \smallskip
\item{11.}S. Weinberg, {\it Phys. Rev.\/} {\bf 166} (1968) 1568.
\smallskip
\item{12.}V. Bernard, N. Kaiser, J. Kambor
and Ulf-G. Mei{\ss}ner, {\it Nucl. Phys.\/} {\bf B388} (1992) 315.
\smallskip
\item{13.}V. Bernard, N. Kaiser, Ulf-G. Mei{\ss}ner and A. Schmidt,
{\it Phys. Lett.\/} {\bf B319} (1993) 269.
\smallskip
\item{14.}M. Benmerrouche, R.M. Davidson and N.C. Mukhopadhyay,
{\it Phys. Rev.\/} {\bf C39} (1989) 2339.
\smallskip
\item{15.}Ulf-G. Mei{\ss}ner, review talk given at WHEPP-III (Third Workshop
on High Energy Particle Physics), Madras, India, Jan. 1994, preprint
CRN--94/04.
\smallskip
\vfill \eject
\noindent{\bf FIGURE CAPTIONS} \bigskip
\item{Fig.1}The process $\gamma (k) + N(p_1) \to \pi^a (q_1) + \pi^b (q_2) +
N(p_2)$. Here, $\gamma$, $\pi$ and $N$ denote the photon, the pion and the
nucleon, in order. $a,b$ are isospin indices, $\epsilon$ is the polarization
vector of the photon and the corresponding four--momenta are indicated.
\medskip
\item{Fig.2}Lowest order diagrams which lead to the leading order result for
the multipoles $M_{1,2,3}$. The circle--cross denotes an insertion from
${\cal L}_{\pi N}^{(2)}$.
\medskip
\item{Fig.3}One--loop diagrams contributing to $M_{1,2,3}$ at order $M_\pi$.
\medskip
\item{Fig.4} (a) Tree diagrams with two insertions
from ${\cal L}_{\pi N}^{(2)}$. (b) Tree diagram sensitive to the anomalous
$\gamma \to 3 \pi$ vertex. These vanish as described in the text.
\medskip
\item{Fig.5}Classification of the
tree diagrams with $\Delta(1232)$ insertions. (a) and (b) lead
to the lowest order ${\cal O}(M_\pi)$ contribution. (c) e.g. contains the
dangerous energy denominator as described in the text. Crossed diagrams are
not shown.
\medskip
\item{Fig.6}(a) Lowest order diagrams contributing to the longitudinal
multipoles $N_{1,2,3}$. (b) The circle--cross denotes an insertion from
${\cal L}_{\pi N}^{(2)}$. The terms proportional to $c_i$ ($i=1,2,3,4$)
and $\kappa_{v,s}$ do not contribute as described in the text.
\medskip
\item{Fig.7}Invariant matrix--elements (in GeV$^{-3}$) for the $\gamma p$
initial state ($k^2 = 0$) as a function
of the off--shell parameter $Z$ (in the isospin limit).
\medskip
\item{Fig.8}Total cross sections (in nb) for the $\gamma p$ initial
state ($k^2 = 0$) as a function of the photon lab energy $E_\gamma$ (in the
isospin limit).
\medskip
\item{Fig.9}Total cross sections (in nb) for the $\gamma p \to \pi^+ \pi^- p$.
The dashed line gives the result from the exact threshold amplitude and the
solid one the first above threshold correction (see appendix A).
\medskip
\item{Fig.10}Total cross sections (in nb) for the $\gamma p$
initial state ($k^2 =0$) with the correct three--body phase space. The
various thresholds are indicated.     \medskip
\item{Fig.11}Tree diagrams with insertions exclusively from ${\cal L}_{\pi
N}^{(1)}$ which lead to the first above threshold correction for
$\gamma p \to \pi^+ \pi^- p$, eq.(A.1).
\medskip
\vfill \eject
{\bf TABLE}
\bigskip \bigskip  \bigskip
$$\hbox{\vbox{\offinterlineskip
\def\strut{\hbox{\vrule height 12pt depth 12pt width 0pt}}
\hrule
\halign{
\strut\vrule# \tabskip 0.1in &
\hfil#\hfil &
\vrule# &
\hfil#\hfil &
\hfil#\hfil &
\hfil#\hfil &
\vrule# \tabskip 0.0in
\cr
&  process &&  ${\cal O}(1)$ & ${\cal O}(1) + {\cal O}(M_\pi)$ &
${\cal O}(1) + {\cal O}(M_\pi)$ & \cr
&          &&                & ${\rm without} \, \, \Delta$    &
${\rm with} \, \,  \Delta$      & \cr
\noalign{\hrule}
& $\gamma p \to \pi^+ \pi^- p$ && 11.01 & 16.10 & 11.22 &\cr
& $\gamma p \to \pi^+ \pi^0 n$ && 7.78  & 14.18 & 20.06 &\cr
& $\gamma p \to \pi^0 \pi^0 p$ && 0.00  & 35.25 & 35.25 &\cr
& $\gamma n \to \pi^+ \pi^- n$ && 11.01 & 12.84 & 10.72 &\cr
& $\gamma n \to \pi^- \pi^0 p$ && 7.78  & 14.18 & 20.06 &\cr
& $\gamma n \to \pi^0 \pi^0 n$ && 0.00  & 30.96 & 30.96 &\cr
\noalign{\hrule}}}}$$
\bigskip
{\noindent\narrower  Table 1:\quad Invariant matrix--elements
$|\eta_1 M_1 + \eta_2 M_2 + \eta_3 M_3 |$
for the various two pion production channels in GeV$^{-3}$ ($k^2 = 0$).
\smallskip}
\vfill \eject \end